\documentclass[12pt,preprint]{aastex}

\def\ltsima{$\; \buildrel < \over \sim \;$}
\def\gtsima{$\; \buildrel > \over \sim \;$}
\def\lsim{\lower.5ex\hbox{\ltsima}}
\def\gsim{\lower.5ex\hbox{\gtsima}}
\def\lapp{\ifmmode\stackrel{<}{_{\sim}}\else$\stackrel{<}{_{\sim}}$\fi}
\def\gapp{\ifmmode\stackrel{>}{_{\sim}}\else$\stackrel{<}{_{\sim}}$\fi}

\newdimen\minuswidth    
\setbox0=\hbox{$-$}
\minuswidth=\wd0
\catcode`@=\active
\def@{\kern\minuswidth}
\setbox0=\hbox{\rm0}
 
\shorttitle{NGC6101} 
\shortauthors{Dalessandro et al.}
 
\begin{document} 
\title{No evidence of mass segregation in the low mass Galactic globular cluster NGC~6101. \footnote{
Based on observations collected at the the Very Large Telescope of the European Southern Observatory, 
Cerro Paranal, Chile (under proposal 091.D-0562). 
Also based on observations with
the NASA/ESA {\it HST} (Prop. 10775), obtained at the Space Telescope Science
Institute, which is operated by AURA, Inc., under NASA contract
NAS5-26555. }  }

\author{
E. Dalessandro\altaffilmark{2}, F. R. Ferraro\altaffilmark{2}, D. Massari\altaffilmark{3,4}, B.
Lanzoni\altaffilmark{2}, P. Miocchi\altaffilmark{2}, G. Beccari\altaffilmark{5}
}
\affil{\altaffilmark{2} Dipartimento di Astronomia, Universit\`a degli Studi
di Bologna, via Ranzani 1, I--40127 Bologna, Italy}
\affil{\altaffilmark{3} INAF- Osservatorio Astronomico di Bologna, via Ranzani 1, 40127 Bologna,
Italy}
\affil{\altaffilmark{4} Kapteyn Astronomical Institute, University of Groningen, PO Box 800, 
9700 AV Groningen, The Netherlands}
\affil{\altaffilmark{5} European Southern Observatory, Karl Schwarzschild Strasse 2, D-85748, Garching bei Munchen,
Germany}

\date{16 July, 2015}

\begin{abstract}
We used a combination of Hubble Space Telescope and ground based data to probe the
dynamical state of the low mass Galactic globular cluster NGC~6101.
We have re-derived the structural parameters of the cluster by using star counts and we find that it 
is about three times more extended than thought before. 
By using three different indicators, namely the radial distribution of 
Blue Straggler Stars, that of 
Main Sequence binaries and 
the luminosity (mass) function, we demonstrated that NGC~6101 shows no evidence of mass segregation, 
even in the innermost regions. 
Indeed, both the BSS and the binary radial distributions fully resemble that of any other cluster population. 
In addition
the slope of the luminosity (mass) functions does not change with  
the distance, as expected for non relaxed stellar systems. NGC~6101 is one of the few globulars 
where 
the absence of mass segregation has been observed so far. 
This result provides additional support to the use of the ``dynamical clock'' calibrated on the radial 
distribution of the Blue Stragglers as a powerful indicator of the cluster dynamical age.
  
\end{abstract}
 
\keywords{binaries: general - blue stragglers; globular clusters: individual (NGC~6101)}

\section{INTRODUCTION}

Globular Clusters (GCs) are the most populous, old and dense stellar aggregates in the Galaxy.
They are formed by millions of stars, whose age, distance and chemical composition can be 
determined with great accuracy. For this reason GCs 
play a crucial role in the current understanding
of stellar and dynamical evolution and they represent the ideal target 
to constrain the interplay between the ``environment'' and the evolution of stars.

The average age of Galactic GCs (GGCs; $<t>\sim 12$ Gyr) is typically significantly larger
than the time-scales in which the internal dynamical processes occur (Meylan \& Heggie 1997).
During their evolution, GCs can survive the early expansion     
triggered by primordial gas expulsion and mass loss due to stellar evolution, then  
their evolution is mainly driven by two-body relaxation
thus reaching higher central
concentrations and eventually the core collapse, while loosing stars through the boundary set by 
the tidal field of their host galaxy
(see for example Heggie \& Hut 2003). 

Two-body relaxation drives the long-term dynamical evolution of GCs.
Because of this physical process, heavier objects tend to sink toward the
cluster centers (mass segregation), while less massive stars are forced towards more
external orbits.
The typical time-scale in which two-body relaxation occurs scales
with the number of stars (Spitzer 1987) and it is typically of the order of 1-2 Gyr in GCs (Meylan \& Heggie 1997).  
The internal dynamics of stellar aggregates affect objects of any mass and it can be efficiently 
probed by means of massive 
test particles, like Blue Straggler Stars (BSSs), binaries and Millisecond Pulsars (e.g. Ferraro et al. 2001, 2003). 
Among them, BSSs 
have been successfully used for this purpose, since they are numerous and relatively easy to measure.
Indeed, Ferraro et al. (2012) have shown that the BSS radial distribution can be 
efficiently used to rank clusters according to their dynamical age.
Using this approach, evidence of mass segregation has been observed in all GCs studied so far,  
with only very few exceptions: $\omega$ Centauri, NGC~2419, Palomar~14, Terzan~8 and Arp~2,  
where the BSS radial distribution has been found to be indistinguishable from that of any 
other population
(Ferraro et al. 2006; Dalessandro et al. 2008b; Beccari et al. 2011; Salinas et al. 2012). 
In general, for clusters that haven't reached relaxation yet, the radial distribution of stars of any mass is
expected to be the same (apart from possible primordial dissimilarities).
The lack of mass segregation in NGC~2419 has been confirmed by Bellazzini et al. (2012)
studying the the radial variation of the luminosity function (LF) of Main Sequence (MS) stars
at different radii\footnote{The same result has been obtained by Baumgardt et al. (2009) who performed 
detailed comparison between velocity dispersion 
profiles and theoretical models.}, while it has been questioned for Palomar~14 by Frank et al. (2014) who observed 
an increase of the slope of the stellar mass function for increasing distance from the cluster center, 
as expected for relaxed systems.

As a part of a large observational campaign aimed at deriving the binary fraction in the external regions
of GGCs, 
we present results about the dynamical state of the low mass ($M_V=-6.94$; Harris 1996- 2010 edition)
GC NGC~6101. This is an old ($t\sim 13$ Gyr, Dotter et al. 2010), metal poor ($[Fe/H]=-1.98$; Carretta et al. 2009) 
Galactic halo GC, with a low concentration ($c=0.80$; Harris 1996). 
A recent analysis of its RRLyrae content (Cohen et al. 2011) revealed that NGC~6101 is an Oosterhoff type II
cluster. This is consistent with the metallicity of the cluster, but it is  unusual 
for its kinematical properties. In fact NGC~6101 is one of the very few metal-poor GCs in 
the Galaxy  with a retrograde motion. Because of its peculiar kinematical properties it has been possibly 
connected
(Martin et al. 2004) to the Canis Major dwarf galaxy. 
The stellar content of NGC~6101 has been studied by 
Sarajedini \& Da Costa (1991) and later on by Marconi et al. (2001; hereafter M01).
In particular M01
performed a detailed analysis of the radial distributions of different stellar populations of NGC~6101.
They found that Horizontal Branch (HB) stars and BSS are more centrally concentrated than MS-Turn Off 
stars. They interpreted
this behavior as evidence of mass segregation among different populations.      

In this study  we investigate the dynamical state of NGC~6101 by using three diagnostics:
the radial distribution of {\it (i)}  BSSs, {\it (ii)} MS binaries,
and {\it (iii)} genuine MS stars with different masses.
According to these three indicators, and at odds with the results of M01, we find that NGC~6101 does not show 
evidence of mass segregation. 
We compare 
this observational fact with theoretical expectations and the new dynamical 
timescale estimates obtained from the newly derived structural parameters.  

The paper is structured as follows: in Section~2 we present the details of our observations and the adopted data
reduction procedures; in Section~3 we derive the main structural parameters of the cluster; in Section~4 the BSS
radial distribution is analyzed and compared to that of the reference populations; in Section~5 we study the binary
content of NGC~6101 at different distances from the center and in Section~6 we analyze the radial variations 
of the luminosity and mass function of MS stars. The main results of the paper are summarized and discussed in
Section~7.

\section{OBSERVATIONS AND DATA ANALYSIS}

The data set used in this paper consists of a combination of images obtained
with both the Hubble Space Telescope (HST) and ground based facilities (see Figure~1).

For the analysis of the BSS radial distribution (see Section~4), we used two publicly available catalogs.
The first is the HST Advanced Camera for Survey/Wide Field Camera (ACS/WFC) catalog 
published by Sarajedini et al. (2007; see also Anderson et al. 2008) in the context of 
the ``ACS Survey of Galactic Globular Clusters''. 
This catalog samples the innermost $\sim120\arcsec$ of the cluster. 
We will refer to this sample as the {\it ACS} data set. The $(V, V-I)$ color magnitude diagram (CMD) 
obtained by using the Johnson-Cousin $V_{\rm ground}$ and $I_{\rm ground}$ magnitudes, is shown in Figure~2.
The second public data set used in this work consists of the B, V and I catalog published and 
fully described by M01 and obtained with the 1.54 mt 
Danish telescope at ESO/La Silla. In the following we will refer to this sample as the 
{\it Danish} sample. The corresponding CMD is shown in Figure~3.

The binary fraction analysis (Section~5) has been performed by using deep $V_{\rm HIGH}$ and 
$I_{\rm BESSEL}$  images obtained with the
the FOcal Reducer/low dispersion Spectrograph 2 (FORS2) at the Very Large Telescope (Prop ID: 091.D-0562; PI: Dalessandro).  
The $2k\times4k$ pixels MIT Red- optimized CCD mosaic in the standard resolution mode 
($\sim0.25\arcsec pixel^{-1}$) was adopted for these observations,
thus to sample the largest possible field of view (FOV; $\sim 6.8\arcmin \times 6.8\arcmin$). 
Three pointings complementary to the {\it ACS} sample have been set up (see Figure~1) starting from $r\sim150\arcsec$ 
(and reaching $r\sim750\arcsec$) from the center of the cluster. 
For each pointing, eight images in the    
$I_{\rm BESSEL}$ band with exposure time $t_{\rm exp}=240$ sec each and four in $V_{\rm HIGH}$
with $t_{\rm exp}=510$ sec, have been obtained. A dither pattern of few arcsec has been adopted 
to allow a better reconstruction of the Point Spread Function (PSF) and avoid CCD blemishes and artifacts.
Master bias and flat-fields have been obtained by using a large number of calibration frames.
Scientific images have been corrected for bias and flat-field by using
standard procedures and tasks contained in the Image Reduction and Analysis Facility 
(IRAF)\footnote{Astronomy Observatory,
which is operated by the Association of Universities fro Research in Astronomy, Inc., under the cooperative 
agreement with the National Science Foundation.}.
The photometric analysis has been performed independently for each image
and chip (see for example Dalessandro et al. 2014) by using \texttt{DAOPHOT{\sc II}} (Stetson 1987).
For each frame we selected several tens of bright, not saturated and relatively isolated stars to model the PSF, 
which turned out to be well reproduced by a Moffat function (Moffat 1969). 
The parameters ($\sigma$, $\beta$) describing the PSF model have been allowed to vary
with a third order polynomial as a function of the instrumental coordinates $(x,y)$ within the frame.
For each chip, the best PSF model was then applied to all sources at $2\sigma$ above the
background by using \texttt{DAOPHOT{\sc II}/ALLSTAR}. We then created a master list of stars composed by sources
detected in at least four frames. In the single frames, at the corresponding positions of the stars present 
in the master-list,
a fit was forced with \texttt{DAOPHOT{\sc II}/ALLFRAME}. 
For each star different magnitude estimates in each filter were homogenized and their weighted mean and standard
deviation were finally adopted as star magnitude and photometric error (see for example Ferraro et al. 1991, 1992).       
We used the stars in common with the Stetson photometric secondary standard catalog (Stetson 2000) to report 
the instrumental magnitudes to the $V$ and $I$ Johnson bands. 
Instrumental coordinates (x, y) have been reported to the absolute ($\alpha$, $\delta$) system by using
the stars in common with the GSC2.3 catalog and the cross-correlation tool \texttt{CataXcorr}\footnote{CataXcorr is
a code aimed at cross-correlating catalogs and finding solutions, developed by P. Montegriffo at INAF -
Osservatorio Astronomico di Bologna, and successfully used by our group for the past 10 years.}.
At this stage the catalogs obtained for each chip are on the same photometric and astrometric system.
They have been combined to form a single catalog that we defined {\it FORS2} sample. 
Stars in common between different pointings have been used
to check for the presence of residuals in the calibration procedure. 

Both the {\it FORS2} and the {\it ACS} catalogs have been corrected for differential reddening 
using the approach described by Milone et al. (2012). We recall the reader to this paper for details on the procedure.
For each star, we used the 50 closest neighbors to compute the corresponding average differential color excess 
$\delta[E(B-V)]$. We adopted extinction coefficients from Cardelli et al. (1989). Over the FOV covered 
by the three pointings, the total variation of $E(B-V)$ amounts to only 0.06 mag. 
The resulting $(V, V-I)$ CMDs are shown in Figure~2.

Since the {\it FORS2} data set saturates at $V\sim18$ and does not provide a complete area coverage (see Figure~1), 
we complemented it with images obtained with the Wide Field Imager (WFI) mounted at the MPG/ESO 2.2 m telescope to
homogeneously analyze the density profile of NGC~6101 (Section~3). 
We used six long exposure images: three in the 
$B_{\rm NEW}$ band with $t_{\rm exp}=120$ sec (Prop ID: 069.D-0582, PI: Ortolani) and three in $V/89_{ESO843}$
with $t_{\rm exp}=60$ sec (Prop ID: 068.D-0265, PI: Ortolani). Both the pre-reduction and the photometric analysis have
been performed as described above for the {\it FORS2} data set. Also the instrumental magnitudes have 
been reported to the Johnson
system by using the stars in common with the catalog by Stetson (2000). A color equation was adopted 
to calibrate the $B_{\rm NEW}$ band, while a zeropoint was enough for the $V/89_{ESO843}$ band.
Instrumental coordinates have been reported to the absolute system for each of the eight WFI chips using the stars 
in common with GSC2.3,
as done before. The resulting $(V, B-V)$ CMD is shown in the right panel of Figure~3.

We emphasize that for all catalogs, the magnitudes are in the Johnson-Cousin photometric system. 
To avoid confusion, in the following we will refer to them as $``B''$, $``V''$ and $``I''$.

\section{DENSITY PROFILE AND CLUSTER PARAMETERS}

We used the {\it ACS} and {\it WFI} data sets to compute the density profile of NGC~6101 from direct star counts.
In particular, the sample is composed of all stars in the {\it ACS} catalog and
those in the complementary {\it WFI} data-set. 
We used stars with $13.5\leq V\leq 19.5$ to avoid incompleteness and saturation problems. 
As done in other works (see for example Dalessandro et al. 2013a), we divided the FOV in 17 concentric annuli 
(five of them are in the {\it ACS} FOV)
centered on the center of gravity (C$_{grav}$), which we have adopted to be the one reported by Goldsbury at al. 
(2010; $R.A.=16^h:25^m:48^s.12$, $Dec.=-72^{\circ}:12\arcmin:07\arcsec.9$).
Each annulus has been then split in an adequate number of sub-sectors (ranging from two to four) 
according to the local density of stars and the angular coverage in the {\it WFI} FOV.
Number counts have been calculated in each subsector and the corresponding densities were obtained dividing them by the
sampled area. Particular attention has been paid to incomplete area coverage of the {\it WFI} data set starting from
$r\sim1000\arcsec$ and to the inter-chip gaps.
The stellar density of each annulus was then defined as the average of the subsector densities, and its standard
deviation was computed from the variance among the subsectors. 
We made sure to guarantee some radial overlap between the two samples in order to homogenize the two portions of the
density profile.
The resulting surface density profile is shown in Figure~4. 
We estimated the contribution of the Galactic field background by averaging the densities of the five outermost measures, 
corresponding to $r>400\arcsec$. We obtain an average background density of $\log(\rho_{\rm bck})\sim-2.7$ stars
arcsec$^{-2}$. We have verified that the observed background density is fully consistent with that obtained 
by using the Besancon Galaxy model simulation (Robin et
al. 2003) covering an area of $1^{\circ}\times 1^{\circ}$ centered on the position of NGC~6101 and for stars in the same 
magnitude limit used to build the density profile. This density has been subtracted to the observed density profile to obtain a ``decontaminated"
density distribution (solid symbols in Figure~4). 
As apparent from the figure, the background in the external regions of NGC~6101 does not appear 
to be constant at odds with what expected, but it mildly increases as a function of the distance. 
This behavior introduces some uncertainties
on the background density estimate. However we have checked that variations of the 
background density within intervals compatible with the observations do not affect significantly 
the density distribution.

We fit the radial density profile by using an isotropic single-mass King model (King 1966) following 
the procedure fully
described in Miocchi et al. (2013). The model best-fitting the observed density profile has a concentration 
$c=1.3^{+0.15} _{-0.16}$ and core radius $r_c=61.3\arcsec ^{+6.9} _{-5.4}$, which yield a tidal radius 
$r_t=1200\arcsec ^{+390} _{-250}$. We also derive the projected half light radius (effective radius)
$r_h=128.2\arcsec ^{+16} _{-7.6}$. 
The newly determined structural parameters are in partial disagreement with those found in
the literature and they make NGC~6101 more concentrated and extended than previously thought. 
In particular, $r_c$ is compatible within the errors with the one listed by Harris 
(1996, $r_c=58.5\arcsec$), while the values of $c$  
(and as a consequence $r_t$) and $r_h$ are much larger than in Harris (1996; $c=0.80$ and $r_h=63\arcsec$).
The same result comes out also from the comparison with the values obtained by McLaughlin \& van der Marel (2005),
who quote $c=0.60$, $r_c=59.2\arcsec$ and $r_h=63.6\arcsec$.
We argue that the difference between our parameters and the literature is mainly due to the fact 
that both Harris
(1996) and McLaughlin \& van der Marel (2005) made use of non homogeneous surface brightness profiles that are limited 
to the innermost $\sim100\arcsec$ from the center not allowing for an appropriate background subtraction.

We have estimated the distance modulus and reddening of NGC~6101 by comparing its CMD to that 
of M~30 (NGC~7099). M~30 can be used as a template for this analysis since it has a similar metallicity 
($[Fe/H]=-2.33\pm0.2$; Carretta et al. 2009) and its distance has been 
robustly determined because of its proximity.
To guarantee homogeneity, we compared the {\it ACS} sample of NGC~6101 with the CMD of M~30 obtained with the same
instrument within the same survey (Sarajedini et al. 2007).

We reported the CMD of M~30 in the absolute plane $M_V, (V-I)_0$ by adopting as distance modulus $(m-M)=14.80$ and reddening
$E(B-V)=0.03$ (Ferraro et al. 2009).
To overlap the CMDs and align all the main evolutionary sequences in the absolute plane (see Figure~5), 
we needed to adopt $(m-M)_V=16.20\pm0.10$ and reddening $E(B-V)=0.12\pm0.02$ for NGC~6101, which yield to a true
(unreddened) distance modulus $(m-M)_0=15.83\pm0.12$ and thus to a distance $d=14.6\pm0.8$ Kpc.
We used these values throughout the paper.
 
These values are slightly larger than those obtained in the literature, but in most cases they are still compatible within the
errors. In particular, Harris (1996) report $(m-M)_V=16.07\pm0.1$ and $E(B-V)=0.05$, 
Sarajedini \& Da Costa (1991) found
$(m-M)_V=16.12\pm0.1$ and $E(B-V)=0.06$, M01 obtained $(m-M)_V=16.12\pm0.03$ $E(B-V)=0.1$, while Cohen et al. (2011) 
estimated $(m-M)_V=16.00\pm0.03$ $E(B-V)=0.1$
from the average luminosity of RRLyrae stars and ZAHB fit.

\section{THE BSS RADIAL DISTRIBUTION}

BSSs in GCs are commonly defined as those stars located along an extrapolation of the MS, 
in a region brighter and bluer than the Turn-Off (TO) point in the optical CMD (Sandage 1962).
Their location suggests that they are more massive than the current cluster population (see for example Ferraro et al.
2006; Lanzoni et al. 2007a). 
Indeed observational evidence
(Shara et al. 1997; Gilliland et al. 1998; Fiorentino et al. 2014) showed that BSSs have masses
of $M\sim 1.2 - 2.0 M_{\odot}$ compared to a mass of $\sim0.8 M_{\odot}$ for stars at the MS-TO of an old stellar system. 

Because of their mass, BSSs are heavily affected by dynamical friction and thus they can be used as natural 
test particles to probe the internal dynamics of stellar aggregates. In particular, their radial distribution has been 
found to probe the efficiency of dynamical friction (Mapelli et al. 2006; Alessandrini et al. 2014; Miocchi et
al. 2015). 
Ferraro et al. (2012) suggested that the BSS radial distribution can be used to define the so called {\it dynamical clock} 
which is a powerful indicator of the cluster
dynamical age. In this picture GCs with a flat BSS radial distribution
are dynamically young stellar systems and they are defined as {\it Family I} GCs, clusters with a 
bimodal radial distribution have intermediate dynamical ages ({\it Family II}), while 
those with a centrally peaked and
monotonically decreasing BSS distribution are dynamically old ({\it Family III}).

The BSS population of NGC~6101 has been studied by Sarajedini \& Da Costa (1991) and M01.
In particular M01 studied also the BSS radial distribution by using a combination of HST Wide Field Planetary Camera 2 
(defined {\it ``hst'' sample}) and ground based data ({\it ``ground'' sample})  
corresponding to the {\it Danish} sample defined in Section~2.
By using cumulative radial distributions and Kolmogorov-Smirnov tests, 
they concluded that BSSs have a large probability (84$\%$) to have 
been extracted from the same parent population as HB, while HB and Tip RGB stars being more 
centrally concentrated than MS-TO stars. They interpreted these results as evidence of mass segregation.

\subsection{Population selections}

We studied the BSS population of NGC~6101 by using the {\it ACS} and {\it Danish} data sets. 
In detail, the final sample consists of stars in the {\it ACS} catalog and those present in the 
{\it Danish} data set and complementary to the {\it ACS} FOV (see Figure~1). 
In this way the central region of NGC~6101 
is homogeneously covered up to a distance $r\sim250\arcsec$ (corresponding to about $4\times r_c$) from 
$C_{grav}$. 
While the {\it WFI} data set
covers a larger FOV and ensures a complete sampling of the cluster up to $r_t$, we preferred this combination of
catalogs for the BSS analysis because of the better photometric quality of the data (see Figures~2 and 3).

As done in previous works (see for example Ferraro et al. 1997, 2004; Lanzoni et al. 2007a,b; 
Dalessandro et al. 2008a and reference therein), 
in order to study the BSS radial distribution, we need a homogeneous selection of BSSs
and at least one reference stellar population. For the case of NGC~6101 we selected HB, RGB and MS-TO
stars as reference. 

BSSs have been selected following the definition of stars bluer and brighter than TO.
The selection box shown in Figure~6 follows the well defined BSS sequence brighter than $V=20$. An additional 
constraint on the color ($(V-I)<0.5$) has been adopted to avoid contamination and blends from MS and SGB stars.
In this way we selected 52 BSSs, 33 in the {\it ACS} FOV and 19 in the complementary {\it Danish} sample.
Because of the more conservative selection criteria adopted in this work, the BSS sample is smaller 
than the one obtained by M01, who identified 73 BSS adopting different magnitude and color selections.
As for the reference populations, we selected RGB stars along the RGB and having $13.5<V<18.7$. 
With these limits we selected a total of 510 RGB
stars, 328 in the {\it ACS} sample and 182 in the {\it Danish} one.
We selected HB stars following their well defined sequence for $(V-I)<0.5$ to minimize the impact
of field contamination. In fact, as apparent from Figure~6, the Galactic field describes a vertical sequence 
in the color range $0.7<(V-I)<0.9$. The HB selection box is shown in Figure~6 and encloses 137 stars. 
By using the list of known RRlyrae stars (Liller 1981; Cohen et al. 2011; Fitzgerald et al. 2012),
we identified 11 stars in common, seven in the {\it ACS} and four in the {\it Danish} FOVs.
The RRLyrae falling outside the HB selection were added to the HB sample.
In this way we count 102 and 42 HB stars in the {\it ACS} and {\it Danish} samples respectively.
In order to provide a direct comparison with M01, we used also MS-TO stars as additional reference population. 
They have been selected approximately within the limits $0.65<(V-I)<0.80$ and $20.2<V<20.9$.
It is worth noting that in this magnitude range, MS stars have a completeness of $100\%$
both in the {\it ACS} (see Section~5.1) and in the complementary data-set (M01).
We count 3125 MS-TO stars in total, 2375 in the {\it ACS} and 750 in the complementary {\it Danish} FOVs.

We obtained an estimate of the Galaxy field contamination by using the Besancon Galaxy 
model simulation (see Section~3). We counted the number of stars lying in the adopted $(V, V-I)$ selection boxes.
We find 125 and 455 Galactic field MS-TO and RGB stars respectively, yielding densities 
$\rho_{bck} ^{MS}\sim 1\times 10^{-4}$ stars arcsec$^{-2}$ and $\rho_{bck} ^{RGB}\sim 3.5\times 10^{-4}$ stars arcsec$^{-2}$. 
We don't find any field star falling in the boxes adopted to select HB and BSS stars.

\subsection{Radial distributions}

We first compared the BSS cumulative radial distribution to 
those of the reference populations. As done in Dalessandro et al. (2013b), we took into account the effect of
Galactic field contamination by statistically decontaminating the RGB and MS populations. We divided the FOV into
four concentric annuli centered on $C_{grav}$ and for each of them we randomly subtracted 
a number of stars corresponding to the average field densities quoted above.
The decontaminated cumulative radial distributions are shown in Figure~7. 
As apparent, the four populations show very similar behaviors.
Indeed when a KS test is applied, we obtain probabilities $P_{\rm BSS/RGB}\sim18\%$, $P_{\rm BSS/HB}\sim20\%$
and $P_{\rm BSS/MS}\sim 65\%$ that they are extracted from a different parent population. 
Since they are well below the canonical probability limit of $95\%$, we can conclude that in all cases 
the differences in the observed radial distributions are not significant. Also, we obtain that the radial distributions of
the parent populations (HB, RGB and MS) are consistent with all being extracted from the same parent distribution. 
These results are in good agreement with what found by M01 as far as the analysis is limited to HB, RGB and BSS stars,
while a different behavior is observed for MS stars.
In fact, M01 found the BSS population to be significantly more centrally segregated than MS stars.
In order to compare more accurately our results with M01,
we used also their {\it ``hst'' sample} and selected stellar populations using the same boxes defined above.
Also in this case, we do not find any significant difference among the radial distribution of RGB, HB and BSS 
and that of MS stars.    

We also analyzed the radial distribution of the specific frequencies $N_{\rm BSS}/N_{\rm RGB}$, 
$N_{\rm BSS}/N_{\rm HB}$ and $N_{\rm BSS}/N_{\rm MS}$, where $N_{\rm POP}$ is the number of stars in the corresponding
population. We divided the FOV into four concentric annuli 
centered on $C_{grav}$ and in each of them we counted the number of BSS and that of the reference population stars.
Galaxy field contamination has been accounted as described above.
The radial distributions of the specific frequencies are shown in Figure~8. In all cases, and in agreement with what 
observed from the cumulative radial distributions,  we find that these ratios are almost constant over
the entire extension of the {\it ACS} + {\it Danish} FOV ($r<250\arcsec$).

We computed also the double normalized ratio ($R_{\rm POP}$\footnote{$R_{\rm POP}=(N_{\rm POP}/N_{\rm POP}^{\rm tot})/(L^{\rm
samp}/L^{\rm samp}_{tot})$, where POP = RGB and BSS. $L^{\rm samp}$ is the luminosity sampled in each annulus and estimated by using
the best-fit King model, distance modulus and reddening derived in Section~3, $L^{\rm samp}_{tot}$ is instead 
the luminosity sampled in the entire field of view.}; Ferraro et al. 1993) for BSS and RGB stars. 
In particular, for the same annuli used before, we estimated the sampled luminosity using the best-fit King model
discussed in Section~3, a distance modulus $(m-M)_{\rm V}=16.20 \pm 0.1$ and reddening
$E(B-V)=0.12$. We find that the $R_{\rm POP}$ is constant for both RGB stars and BSSs  
as expected for any post-MS population (Renzini \& Fusi Pecci 1988).
The same result is obtained when HB and MS stars are considered.

Although this analysis does not cover the entire extension of NGC~6101 (it is limited to $\sim2 r_h$),
we note that the considered data-sets sample about 
the $70\%$ of the total luminosity of the cluster. On the basis of what observed 
in tens of GCs (Ferraro et al. 2012), we do not expect any significant deviation
from the general behavior at larger distances. 
Therefore we can safely conclude that the BSS radial distribution of NGC~6101 is flat (i.e., indistinguishable from that of the
reference populations). Such a distribution
is not common for GCs and it has been observed only in few other cases so far ($\omega$
Centauri - Ferraro et al. 2006; NGC~2419 - Dalessandro et al. 2008b; Palomar~14 - Beccari et al. 2011; Ter~8 and Arp~2
- Salinas et al. 2012).
As for those GCs, a flat BSS radial distribution is a strong indication of the absence of 
mass segregation in NGC~6101.

\section{THE BINARY FRACTION RADIAL DISTRIBUTION}

The fraction of binaries is an essential component of the formation and evolutionary processes of stellar systems.
In dynamically active aggregates as GCs, binaries are thought to promote the formation of 
exotic objects like BSSs, X-ray sources and Millisecond Pulsars (see e.g. Mc Crea et al. 1964,
Paresce et al. 1992; Ferraro et al. 2006b, 2009; Heinke et al. 2003; Pooley \& Hut 2006; Xin et al. 2015).
Binaries are typically more massive than the average single star mass in GCs ($<m>\sim0.3 M_{\odot}$), 
therefore they tend to sink towards the center of GCs
because of dynamical friction. As a consequence, also their radial distribution can be an useful tool to constrain the
dynamical state of GCs.

Up to now, three main techniques have been used to measure the binary fraction:
{\it (i)} radial velocity variability (e.g. Mathieu \& Geller 2009), {\it (ii)} search for eclipsing binaries 
(e.g. Mateo 1996; Cote et al. 1996), and {\it (iii)} studies of the distribution of stars along the cluster MS in CMDs 
(Romani \& Weinberg 1991; Bellazzini et al. 2002).  
In this paper we used the latter approach by following the method described by Bellazzini et al. (2002; 
see also Sollima et al. 2007; Dalessandro et al. 2011; Beccari et al. 2013). 
The basic idea is that the magnitude of the binary system corresponds 
to the luminosity of the primary star (more massive) increased by that of the companion.
Stars on the MS obey a mass-luminosity relation, hence the luminosity of the binary system is a function of the 
mass ratio $q=m_2/m_1$ of the two components (where $m_1$ and $m_2$ are the masses of the primary and secondary, respectively). 
Since $q$ can assume any value between $0$ and $1$, in CMDs binaries describe a ``secondary MS'' in the CMD, i.e. they generate
a broadening of the single stars MS, toward higher luminosities.

The binary fraction ($\xi$) of the central regions of NGC~6101 has been estimated by Sollima et al. 
(2007) and  Milone et al. (2012). 
In this work, we recomputed the central binary fraction by using the {\it ACS} sample for homogeneity, 
and we extended the analysis to the cluster peripheries by using the {\it FORS2} data set.

\subsection{Artificial stars experiment and Galaxy contamination.}

In order to estimate the binary fraction by analyzing the ``secondary MS'', a number of spurious effects should be considered.
First, the Galactic field contamination should be properly taken into account.
Then, a robust measures of source of broadening, like blending and photometric errors, should be considered.
These factors are related to the quality of the data and can be properly studied through artificial stars experiments.

For the {\it ACS} sample, we used the artificial star catalog provided with the real stars catalog and available at the
``ACS Survey of Galactic Globular Clusters'' web page\footnote{http://www.astro.ufl.edu/$\sim$ata/public
$\textunderscore$hstgc/}. For the {\it FORS2} sample    
we performed a large number of artificial stars experiments. We followed
the method described by Bellazzini et al. (2002; see also Dalessandro et al. 2011). 
We generated a catalog of simulated stars with an I-band input magnitude ($I_{\rm in}$) magnitude extracted from a 
LF modeled to reproduce the
observed luminosity function (LF) in that band and extrapolated beyond the limiting magnitude. Then to each star
extracted from the LF we assigned a $V_{\rm in}$ magnitude by means of an interpolation along the mean ridge line of the
cluster.

Artificial stars were added to real images by using the \texttt{DAOPHOT{\sc II}/ADDSTAR} software. In order to avoid 
``artificial crowding'', stars were placed into the images following a regular grid composed by $35\times35$ pixels 
cells
(corresponding approximately to 8-9 times the usual FWHM of stars for these images) in which only one artificial star for
each run was allowed to lie.  More than 250,000 stars have been simulated for the entire FOV covered by the {\it FORS2} sample.
The photometric reduction process used for the artificial star experiments is exactly the same as described in
Section~2. Those stars recovered after the photometric analysis have also values of $V_{\rm out}$
and $I_{\rm out}$. 
As already noticed in Dalessandro et al. (2011; see also Milone et al. 2012), the MSs of the artificial stars
CMDs are narrower than
the observed ones. One possible reason for this effect, is that the formal photometric errors of the artificial-star catalogs
systematically underestimate the true observational uncertanties. Another possibility is that the broadening of the 
observed data is due to some physical effects, as the presence of multiple populations along the MS. 
Irrespective of the origin and as done in Dalessandro et al. (2011) we increased the formal artificial-star catalog errors thus to
reproduce the observed error distribution as a function of magnitude.
The curves of photometric completeness ($C$), defined as the ratio between the number of stars recovered at the
end of the procedure and the total number of stars actually simulated, are shown in Figure~9 for different radial bins.

For a proper measurement of the binary fraction we performed a detailed study of the field contamination,
using the same 
Besancon Galaxy model catalog described in Section~4.

\subsection{The binary fractions}

The high photometric quality and the spatial coverage provided by the combination of the {\it ACS} and {\it FORS2} data sets, 
allow us to study the binary fraction radial distribution from the core to up $600\arcsec$ from the cluster center.\footnote{
While the {\it FORS2} data set extends out to $r\sim800\arcsec$, we preferred to limit the analysis at $r\sim600\arcsec$ where the fraction of
covered area coverage is more complete and the number of detected stars is appropriate for an accurate analysis of the binary population.}
We divided the FOV in four concentric annuli roughly corresponding to $0.0-1.0$ $r_h$, $1.0-2.0$ $r_h$, $2.0-3.5$ $r_h$ and 
$3.5-5 r_h$, and we estimated both the minimum ($\xi_{\rm min}$)  and 
the global binary fractions ($\xi_{\rm TOT}$) following the approach described by Bellazzini et al. (2002). 
The analysis was performed for stars with $19\leq I\leq 22$, where the completeness parameter ($C$) 
is safely larger than $50\%$. 
This interval corresponds to a mass range $0.55<m/M_{\odot}<0.80$ for a single star along the MS, according to 
an isochrone of 12 Gyr and $Z=0.003$ (Bressan et al. 2012).

First, we measured $\xi_{\rm min}$, which is the fraction of binaries with a mass ratio $q_{\rm min}$
large enough to make them clearly distinguishable from single MS stars. The value of $q_{\rm min}$ depends on
the photometric errors: we considered a color range equal to  
three times the photometric error 
from the MS ridge line (see Figure~3 in Dalessandro et al. 2011); 
in the considered magnitude interval and for the entire sample such a color difference corresponds to $q_{\rm min}\sim0.4$.
The contamination from blended sources and non member stars has been accounted by means of the artificial star 
catalogs and the Besancon model simulation (Section~5.1).
The radial distribution of $\xi_{\rm min}$ is shown in Figure~10. As already observed for BSSs, binaries 
show no significant evidence of mass segregation. In fact, their radial distribution 
is almost flat around a mean value $\xi_{\rm min}=(6.2\pm 0.4)\%$.
Such a value is slightly larger than what obtained by Milone et al. (2012) only in the central regions 
($q_{\rm min}=4.8\pm 0.3\%$). It is important to note, however, that Milone et al. (2012) 
considered only binaries with a mass ratio larger than ours ($q>0.5$).

As a second step of the analysis, we estimated also $\xi_{\rm TOT}$. 
To this aim we created hundreds of synthetic CMDs including 
single and binary stars for different input values of the global binary fraction 
($\xi_{\rm IN}$), for each run.
To simulate the binaries we randomly extracted $N_{\rm bin}$ values of the mass of the primary from a Kroupa (2002)
initial mass
function and $N_{\rm bin}$  values of the mass of the primary component from the Fisher et al. (2005) 
mass ratio distribution, $f(q)$, which is one of the unknowns in this kind of analysis.
The impact of adopting different
mass ratio distributions has been discussed in Sollima et al. (2007) and Dalessandro et al. (2011). 
To generate the sample of single MS stars we adopted the best fit to the observed mass function (MF),
which is discussed in detail in Section~6.
In Dalessandro et al. (2011) we have verified that the adoption of different present day
mass functions to simulate MS stars has a negligible impact on the final result.
$\xi_{\rm TOT}$ was then determined from the comparison between the artificial and the observed CMDs: 
the value of $\xi_{\rm IN}$ providing the best match between the two CMDs is adopted as global
binary fraction $\xi_{\rm TOT}$. In particular, from the simulated catalog we computed the ratio 
$r_{\rm sim} = N_{\rm bin}^{sim}/N_{\rm MS}^{sim}$ between the number of
binary stars defined as before and that of the synthetic MS population. The same was done for the observed CMD, 
thus obtaining $r_{\rm sim} = N_{\rm bin}^{obs}/ N_{\rm MS}^{obs}$. For different values of $\xi_{IN}$ 
we computed the penalty function $\chi^2(\xi_{IN})$
defined as the summation of $(r_{\rm sim}-r_{\rm obs})^2$ and the relative probability $P(\chi^2)$ was derived (Figure~11).
The mean of the best-fitting Gaussian is adopted ad best fit value of the global binary fraction $\xi_{\rm TOT}$. 
The radial distribution of $\xi_{\rm TOT}$ is shown in Figure~10. 
Similarly to the minimum binary fraction, also $\xi_{\rm TOT}$ shows no variation as a function of 
the distance from the center in the FOV covered in this analysis.
This results confirms that NGC~6101 has not experienced complete relaxation yet and does not show any evidence of mass segregation
among its populations.

The mean value of the global binary fraction is $\xi_{\rm TOT}=(14.4\pm 0.9)\%$. 
This value is in good agreement with the one obtained by Sollima et al. (2007) in the {\it ACS} FOV 
($\xi_{\rm TOT}=15.6\pm 1.3\%$), while it is significantly larger than the one estimated by 
Milone et al. (2012; $\xi_{\rm TOT}=9.6\pm0.6 \%)$. Note however that Milone et al. (2012) determined the value of 
$\xi_{\rm TOT}$ assuming that it is twice the value of $\xi_{\rm min}$.

\section{LUMINOSITY AND MASS FUNCTION RADIAL VARIATIONS}

To further investigate the mass segregation phenomenon in NGC~6101,
we analyzed the radial
variation of the luminosity function (LF) of MS stars. LFs give information about the effect of cluster internal 
dynamics on stars in a wide range of masses, including the faint-end of the MS where most of the cluster mass lies.
In relaxed stellar systems, the slope of the LFs is expected to vary as a function of the distance from the cluster center,
with indexes decreasing as the distance increases, because of the differential effect of mass segregation.
 
To study the LF of NGC~6101 we used the {\it ACS} and {\it FORS2} samples. We selected a sample of bona-fide stars along the MS, 
defined as those stars located within $3\sigma$ from the mean ridge line, 
where $\sigma$ is the combined photometric uncertainty in the $V$ and $I$ bands.   
We used the same radial bins defined in Section~5 for the binary fraction analysis. 
For each radial interval we obtained the completeness-corrected
LF for stars fainter than $V=18$ and reaching the magnitude limit where
the completeness parameter $C$ is $50\%$.
Completeness has been derived by means of the artificial star catalog described in Section~5.1 (see Figure~9).
The LFs obtained in the four radial bins are shown in Figure~12.
We took the LF corresponding to the innermost radial bin as reference for comparison with the LFs in the other radial ranges. 
The line connecting the points of the reference LF is plotted as a solid line in Figure~12. As apparent, 
it  well reproduces the behavior observed in the more external radial bins after a proper normalization. 
The key result of Figure~12 is that, in the range of magnitude and radial extension ($0<r<5 r_h$) covered by our analysis, 
the shape of the different LF is clearly the same and any subtle
difference lies well within the combined uncertainty of the LFs and of the adopted normalization. 

We derived the mass function (MF) of the same bona-fide stars used for the LF and within the same magnitude limits.
Masses have been obtained by using the Baraffe et al. (1997) mass-luminosity relation.
The MFs obtained in the same radial bins as the LFs are shown in Figure~13. The MF obtained in the {\it ACS} sample
goes down to $m\sim0.15 M_{\odot}$, while it reaches at most $m\sim0.35 M_{\odot}$ in the {\it FORS2} data set. 
As done before, we used the MF of the innermost radial bin as reference, and we normalized the more external ones
to this by using number counts in the mass range $(0.5 <m/M_{\odot}<0.7)$. As apparent from Figure~13, 
and in agreement with what obtained from the analysis of
the LFs, all MFs are virtually indistinguishable.

\section{DISCUSSION}

Three distinct indicators, namely the radial distribution of BSSs, binaries stars and 
the LF of MS stars have been investigated to study the mass segregation in NGC~6101.
We have traced this effect by using stars in a wide range of masses, going from $\sim0.35 M_{\odot}$, corresponding
to the lowest mass MS stars, up to $\sim1.4M_{\odot}$ reached by the highest mass BSSs and binaries. 
The results clearly indicate that the cluster  do not show evidence of mass segregation 
up to $r\sim600\arcsec$ ($\sim5r_h$) from the cluster center.

Only few other GGCs are known to be in such a dynamical state:
massive systems ($M_{\rm V}<-10$), namely $\omega$~Centauri (Ferraro et al. 2006a) and
NGC~2419 (Dalessandro et al. 2008b), and three GCs that are at the low mass end distribution ($M_{\rm V}\sim-5$),
namely Palomar~14 (Beccari et al. 2011), Ter~8 and Arp~2 (Salinas et al. 2012).
Similarly to these clusters, NGC~6101 can be considered dynamically young.

Starting from the new structural parameters obtained in Section~3, 
we derived the central ($t_{\rm rc}$) and half mass radius ($t_{\rm rh}$)
relaxation times following the formulae by Spitzer (1987). We adopted the same values of
$(m-M)_V$ and $E(B-V)$ used in Section~4.2, 
 which yield a distance from the Sun $d=14.6\pm 0.8$ Kpc (see Section 3). We estimated the total 
cluster mass $M_t$ by adopting an integrated V magnitude $V_t=9.2$ (Harris 1996) and a
mass-to-light ratios $M/L_V=3.17$ and 2.07 (Maraston et al. 1998), 
appropriate for a Salpeter and Kroupa IMF respectively, and for a population of metallicity $[Fe/H]=-2.25$ 
and age $t=13$ Gyr.
We obtained $M_t=1.2-1.7\times10^5 M_{\odot}$ which gives a central mass density 
$\rho_{M,0}=68.5 M_{\odot}$ pc$^{-3}$. We thus obtained $t_{rc}\sim1.3$ Gyr and $t_{rh}\sim5.4-6.3$ Gyr. 
The value of $t_{rc}$ is consistent with what reported 
by Harris (1996), while $t_{rh}$ is about three times larger than the Harris value ($t_{rh}\sim1.7$ Gyr) and more 
than five times larger than the one
derived by McLaughlin \& van der Marel 2005 ($t_{rh}\sim0.9$ Gyr). This is qualitatively in agreement with  
$r_h$ being larger than previous estimates reported in the literature.

We note however that both $t_{rc}$ and $t_{rh}$ are significantly smaller than the age of NGC~6101
($t_{\rm age}=13\pm 1$ Gyr; Dotter et al. 2010).   
Therefore some evidence of mass segregation should be visible, at odds with the observed flat
distribution of BSS and binaries, and the constant slope of the MF LFs.
In the cases of NGC~2419 and Palomar~14, $t_{\rm rh}$ has been estimated to be $\sim 20$ Gyr 
(Dalessandro et al. 2008b; Sollima et al. 2011), in good agreement with the observational evidence 
of a lack of mass segregation, at least in the external regions. 
However also for NGC~2419,
the central relaxation time is significantly smaller ($t_{\rm rc}\sim6$ Gyr) than the age of the cluster ($t=12$ Gyr)
and Ferraro et al. (2006) found for $\omega$~Centauri 
a relatively short relaxation time ($t_{\rm rh}\sim 5$ Gyr), as for NGC~6101.
In the case of $\omega$~Centauri a number of possibilities to understand this apparent disagreement between 
observations and theoretical results were examined. 
Cluster rotation can play a role. In fact angular momentum
tends to keep stars out of the core, balancing the effect of mass segregation (Spurzem 2001). 
Another possibility is that $\omega$ Centauri could have been hundreds of time more massive in the past 
than observed today. Indeed 
it has been suggested to be the relic of a partially disrupted galaxy (Bekki \& Freeman 2003; Tsuchiya et al. 2004).
The same argument can be applied to
NGC~2419 (van den Bergh \& Mackey 2004; Mackey \& van den Bergh 2005) and the extended tidal tails observed around Palomar~14 (Sollima et al. 2011) would suggest that 
it experienced quite recent and efficient tidal stripping events.
  
For the case of NGC~6101 no evidence of internal rotation are known. 
However we may speculate that NGC~6101 could have experienced a quite complex dynamical 
history. 
By means of dynamical simulations, 
Martin et al. (2004) suggested that NGC~6101 can be associated to the Canis Major dwarf galaxy 
accreted to the Galactic halo after an encounter between the dwarf and the dark matter halo of the Milky Way.
Indeed the kinematical properties of  NGC~6101 are atypical for systems with its metallicity and age. 
In fact it is one of the very few metal-poor and old GGCs 
with a retrograde motion (Geisler et al. 1995; Rutledge et al. 1997).
Mackey \& Gilmore (2004) suggested that the existence of old GCs with large cores in a 
relative proximity to the Galactic center (as for the case of NGC~6101) can be explained by the fact that they follow 
wide orbits in which they spend little time close to the Galactic center. 
Alternatively, they could have been accreted only recently by the Galactic halo.

In this respect it is interesting to note that a linear fit to the MFs in Figure~12 limited to the stars 
with $m<0.75 M_{\odot}$ gives a power-law index $\alpha\sim-0.9$. 
With such a value and with the new estimate of the central concentration obtained in Section~3 ($c=1.3$),
NGC~6101 would nicely fit the correlation found by De Marchi, Paresce \& Pulone (2007) according to which less concentrated GCs 
tend to have flatter MFs. 
This behavior was interpreted by the authors as an indication of a more efficient loss of low mass stars
via evaporation and tidal stripping in less concentrated clusters. To support this possibility, evidence of tidal arms and 
distortions should be searched for this cluster.

On the other hand, the observational facts collected in the case of NGC~2419, $\omega$ Cen and NGC~6101
suggest that the current theoretical estimates of the central relaxation time are extremely rough and must be used (at
least in absolute terms) with caution. 
On the other side the nice agreement among the different mass segregation indicators used in the case of NGC~6101
provides additional support to the use of the radial distribution of BSS (the so called ``dynamical clock'')
as a powerful indicator of the dynamical evolution of stellar systems (Ferraro et al. 2012).
In fact, among the adopted mass-segregation indicators, BSSs are significantly brighter than the others (MS stars and 
binaries), their analysis is simpler and less prone to observational bias (as completeness) and assumptions.
Hence the BSS radial distribution represents the clearest indicator
of mass segregation in stellar systems.

\acknowledgements 
This research is part of the project COSMIC-LAB (http://www.cosmic-lab.eu) funded by the European Research Council
(under contract ERC-2010-AdG-267675).

\newpage

\begin{figure}
\includegraphics[scale=0.7]{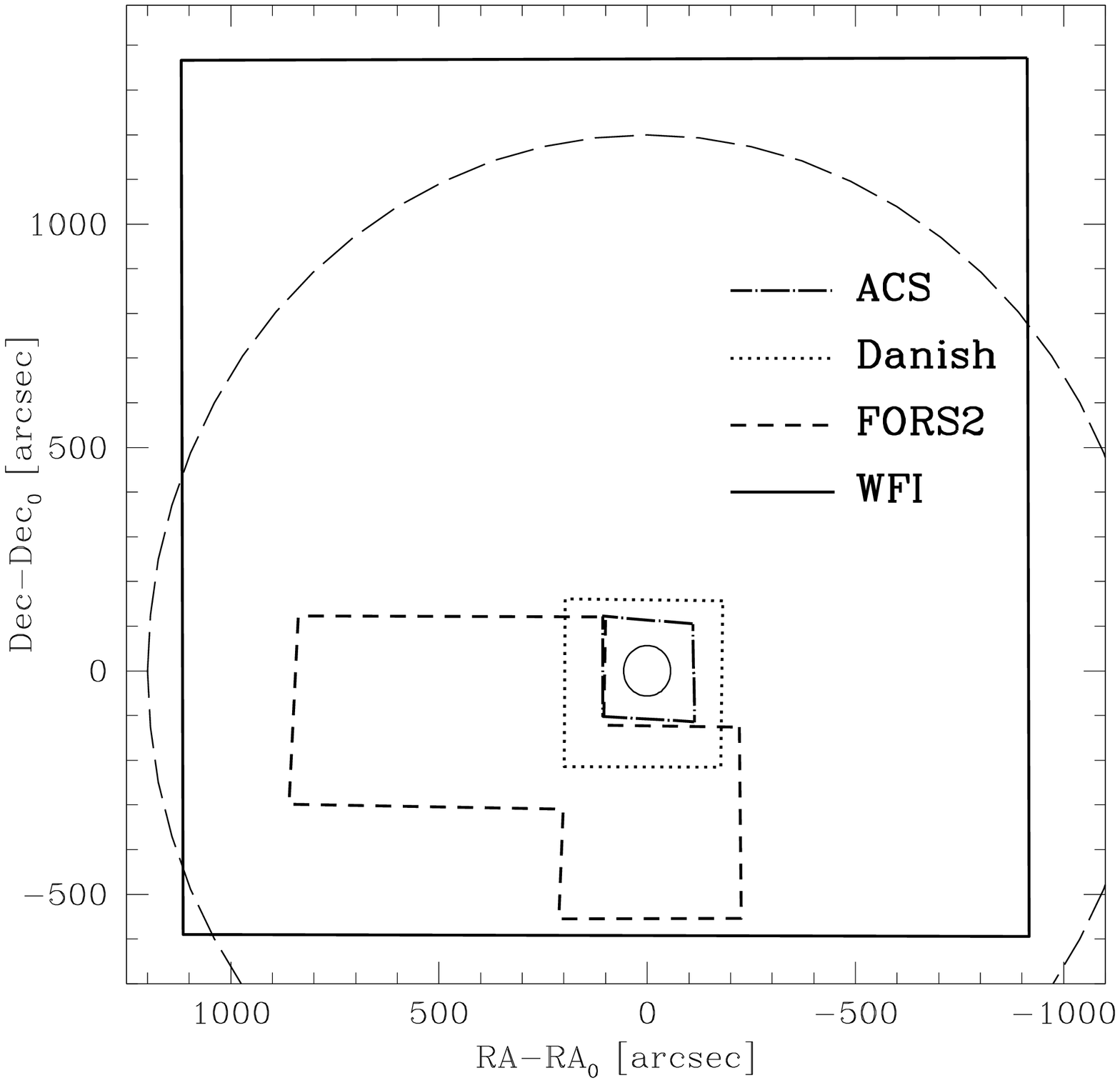}
\caption{Schematic map of the entire database used in this work with respect to the position 
of the center of the cluster. The solid and dashed circles represent the clusetr core and tidal radii respectively.}
\end{figure}

\begin{figure}
\includegraphics[scale=0.7]{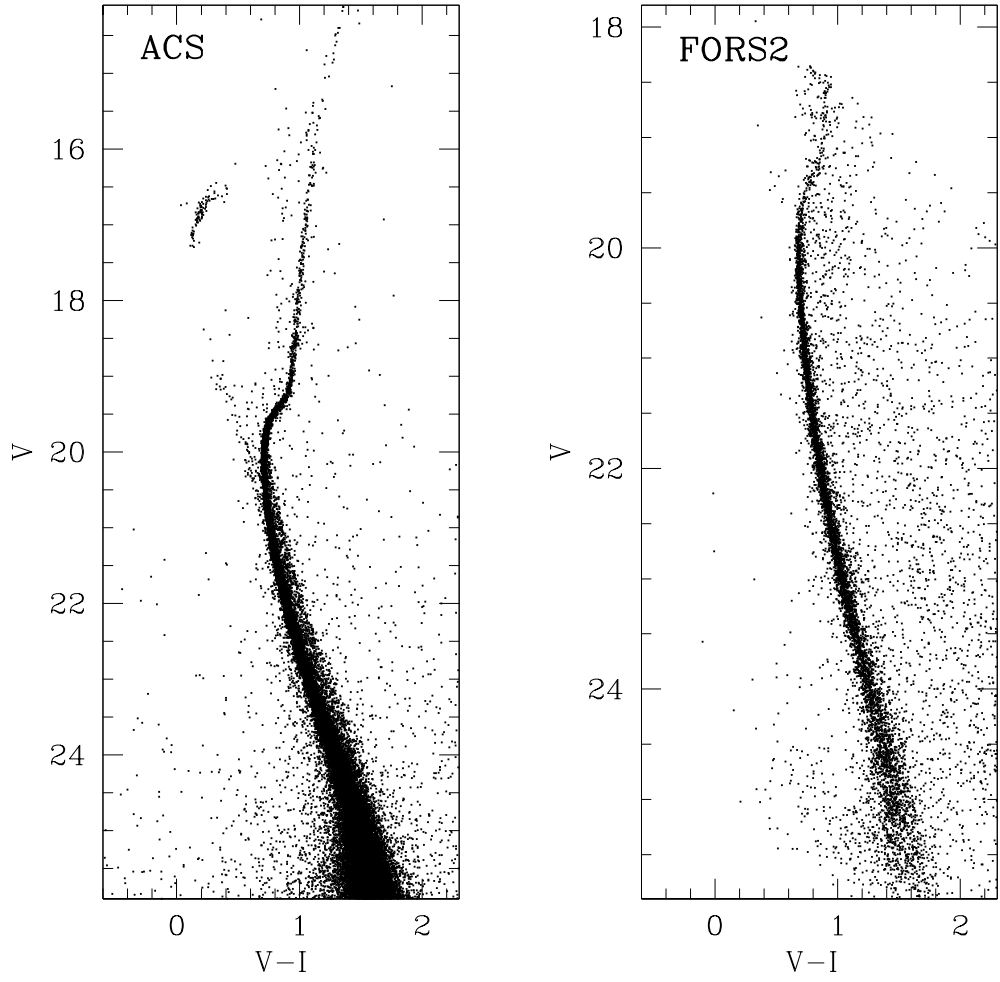}
\caption{$(V, V-I)$ CMDs of the ACS and FORS2 data sets (Section~2). }
\end{figure}

\begin{figure}
\includegraphics[scale=0.7]{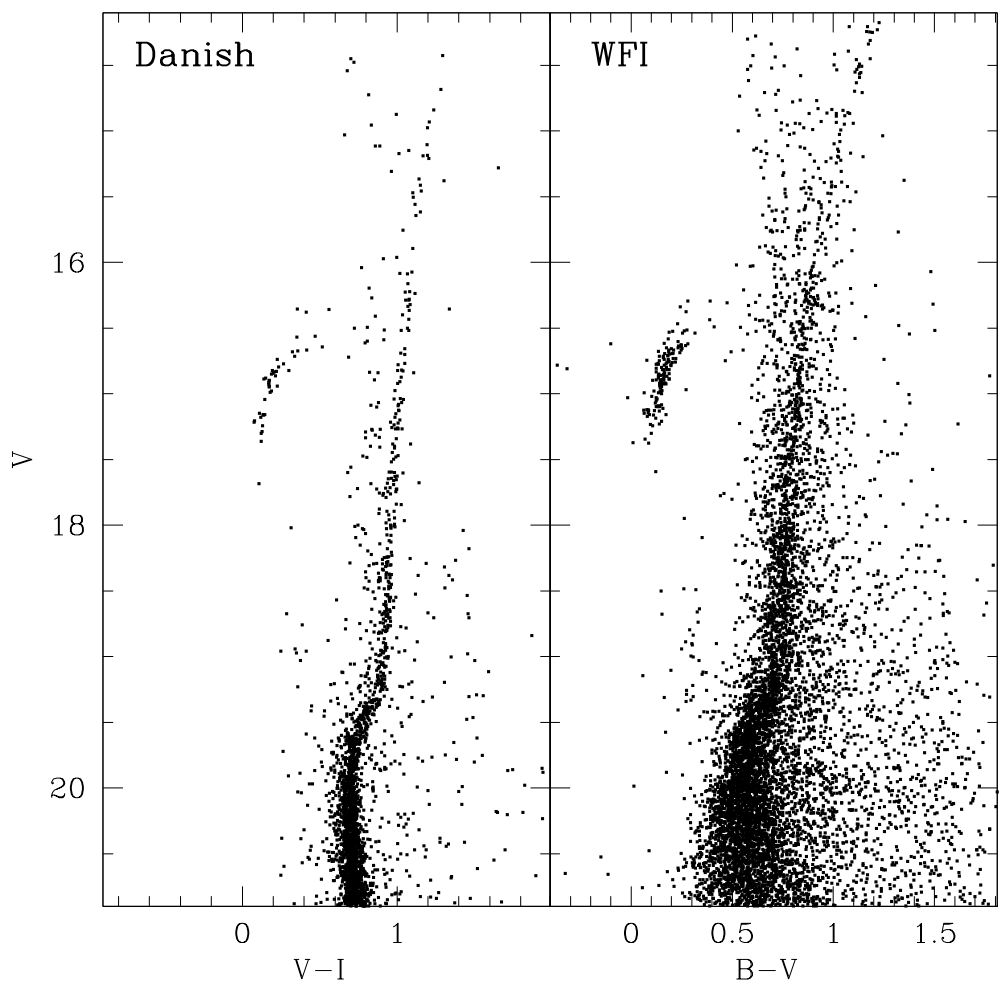}
\caption{{\it Left panel}: $(V, V-I)$ CMD of the {\it Danish} data set published by Marconi et
al. (2001). Here only the stars lying outside the ACS FOV are shown. {\it Right panel}: 
$(V, B-V)$ CMD obtained from the WFI sample. In order to minimise the contamination from Galactic
field stars, we show here only stars located at $r<500\arcsec$ from the cluster center.}
\end{figure}

\begin{figure}
\includegraphics[scale=0.7]{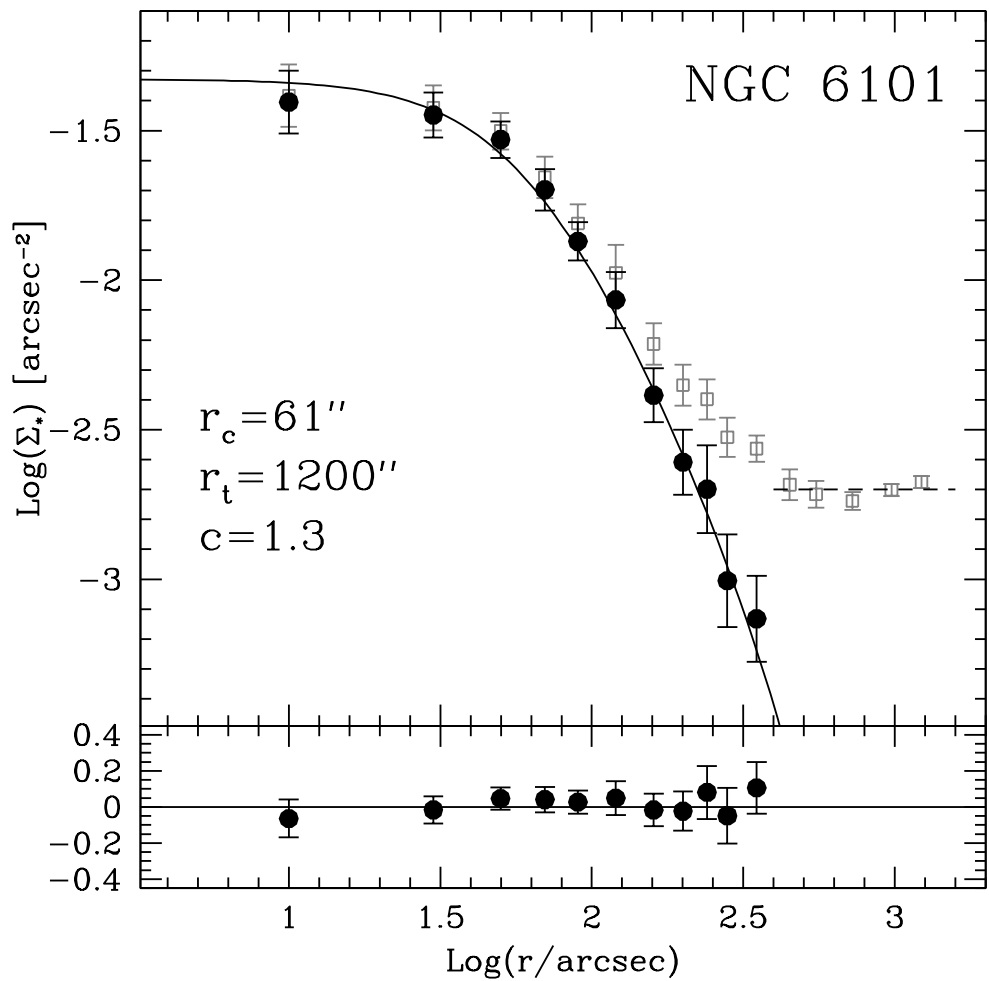}
\caption{Observed star count density profile as a function of radius (open squares). The dashed line
represent the density value of the Galactic field background, obtained averaging the five outermost points. 
The black filled dots are densities obtained after background subtraction (Secion~3). The best-fit
single-mass King model is also overplotted to the observations (solid line). 
The structural parameters are labeled. The lower panel shows the residuals between the observations
and the best-fit model.}
\end{figure}

\begin{figure}
\includegraphics[scale=0.7]{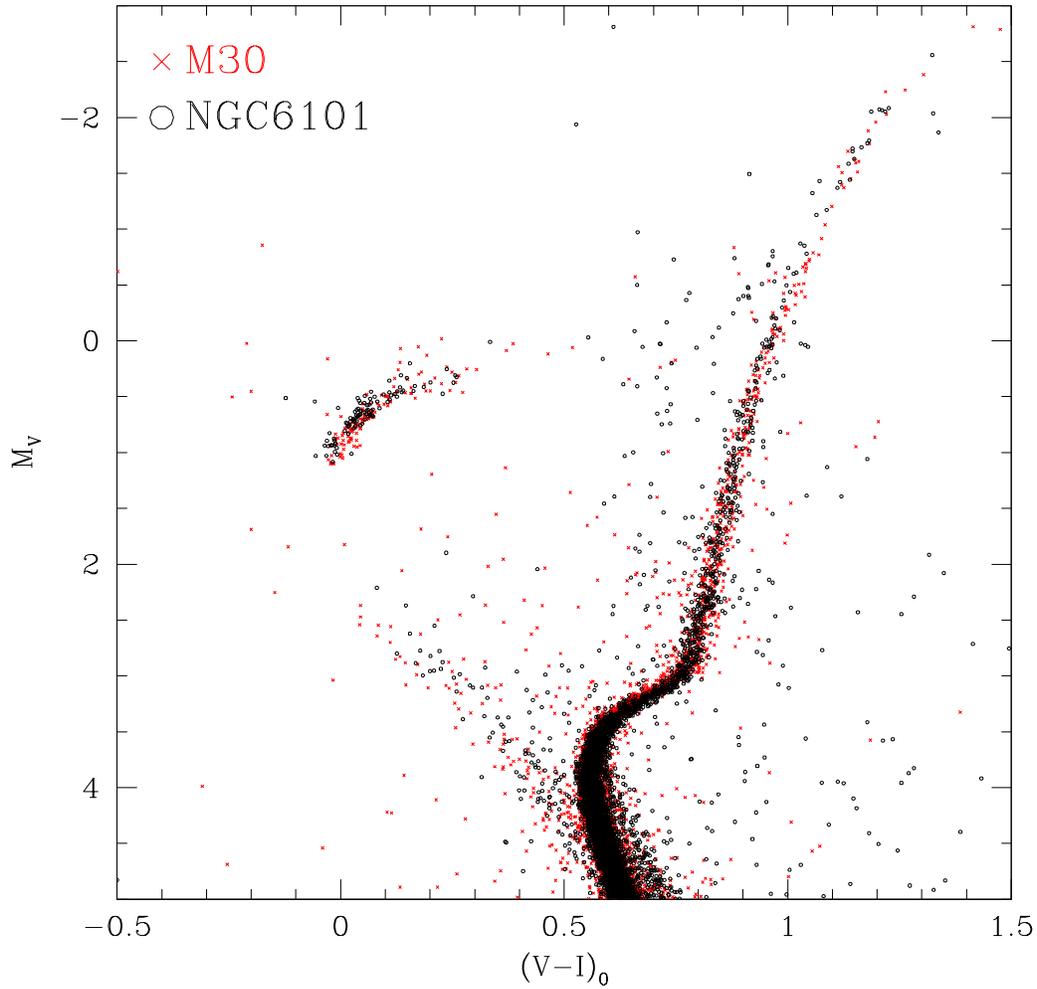}
\caption{Superposition of the CMD of NGC~6101 (black open circles) to that of M~30 (red crosses) 
in the absolute $M_V, (V-I)_0$ plane. For M~30 we adopted the values of distance and extinction obtained by Ferraro et
al. (2009). A distance modulus $(m-M)_V=16.20\pm0.1$ and a color excess $E(B-V)=0.12$ were then adopted 
for NGC~6101 to match the main evolutionary sequences of M~30.}
\end{figure}

\begin{figure}
\includegraphics[scale=0.7]{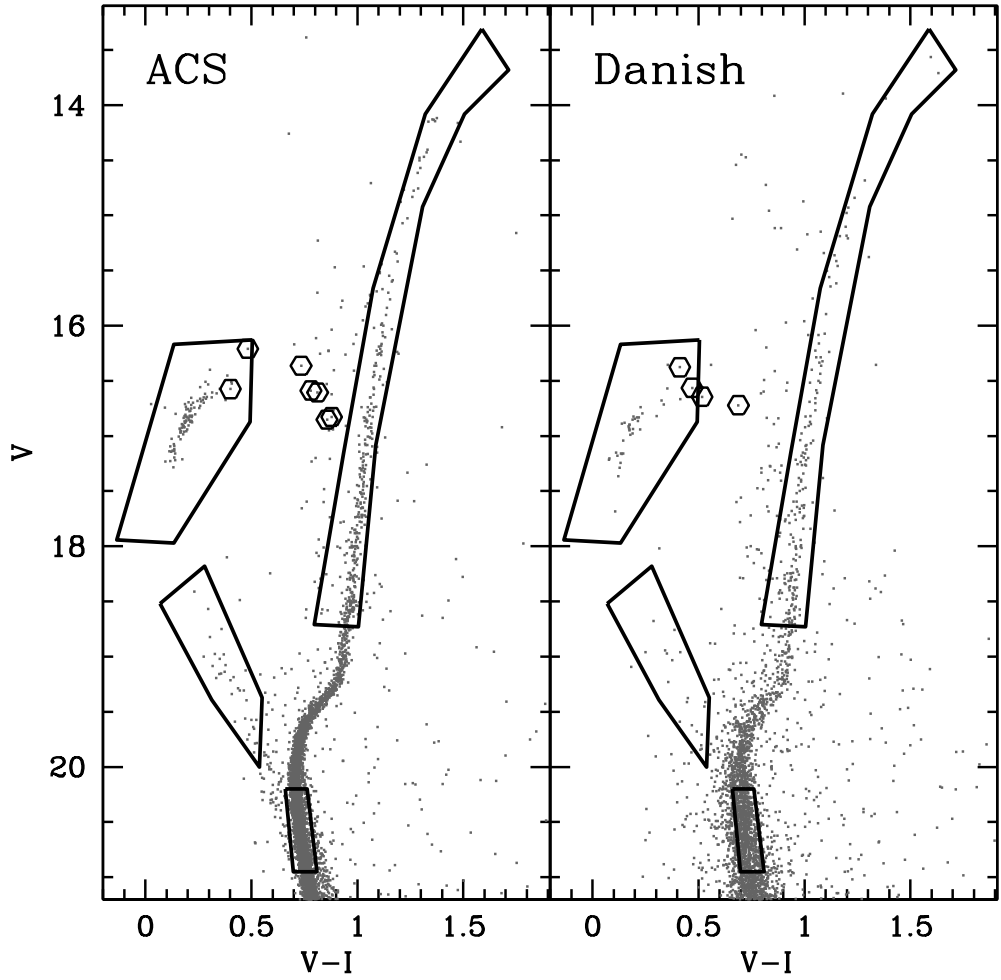}
\caption{Zoomed view of the $(V, V-I)$ ACS and ground based CMDs. Selection boxes for MS, RGB, HB stars and
BSSs stars are shown. The RRLyrae identified in the HB sample are marked with open exsagons. }
\end{figure}

\begin{figure}
\includegraphics[scale=0.7]{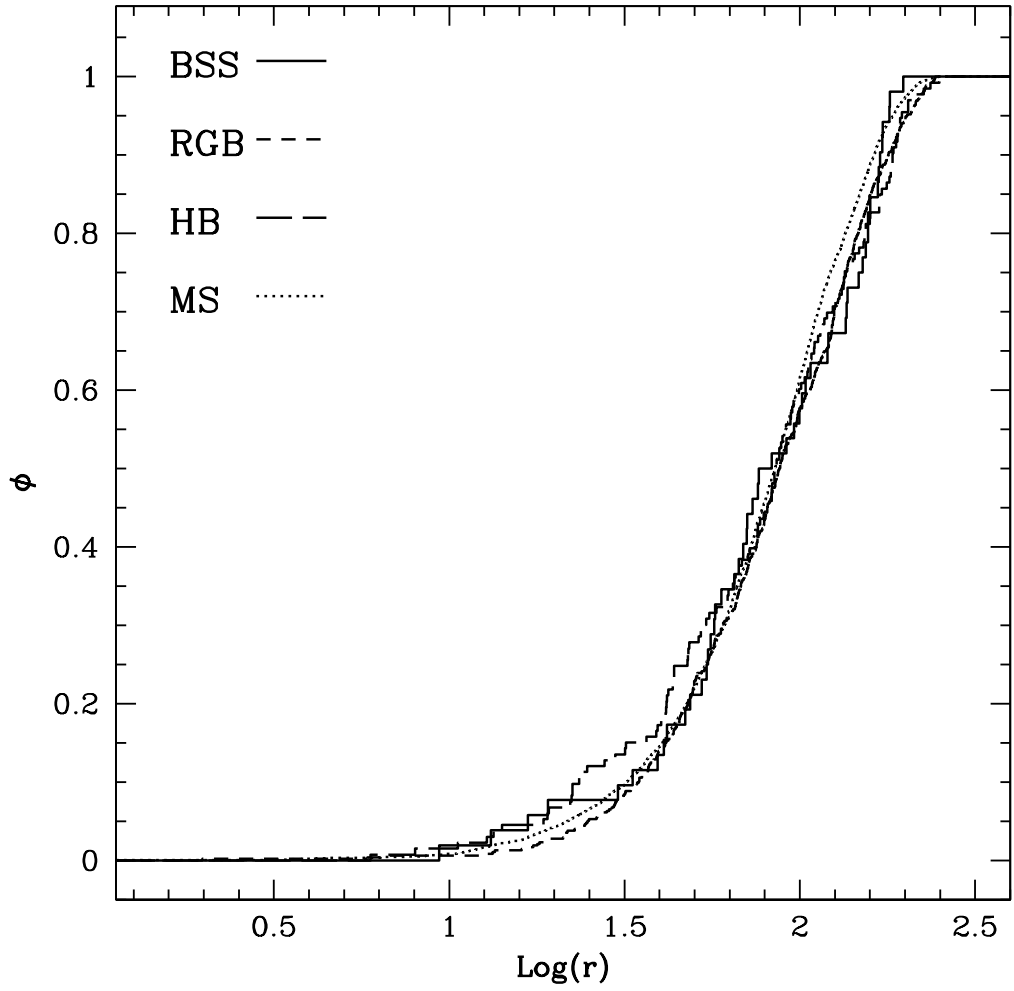}
\caption{Cumulative radial distribution of the statistically decontaminated populations.}
\end{figure}

\begin{figure}
\includegraphics[scale=0.7]{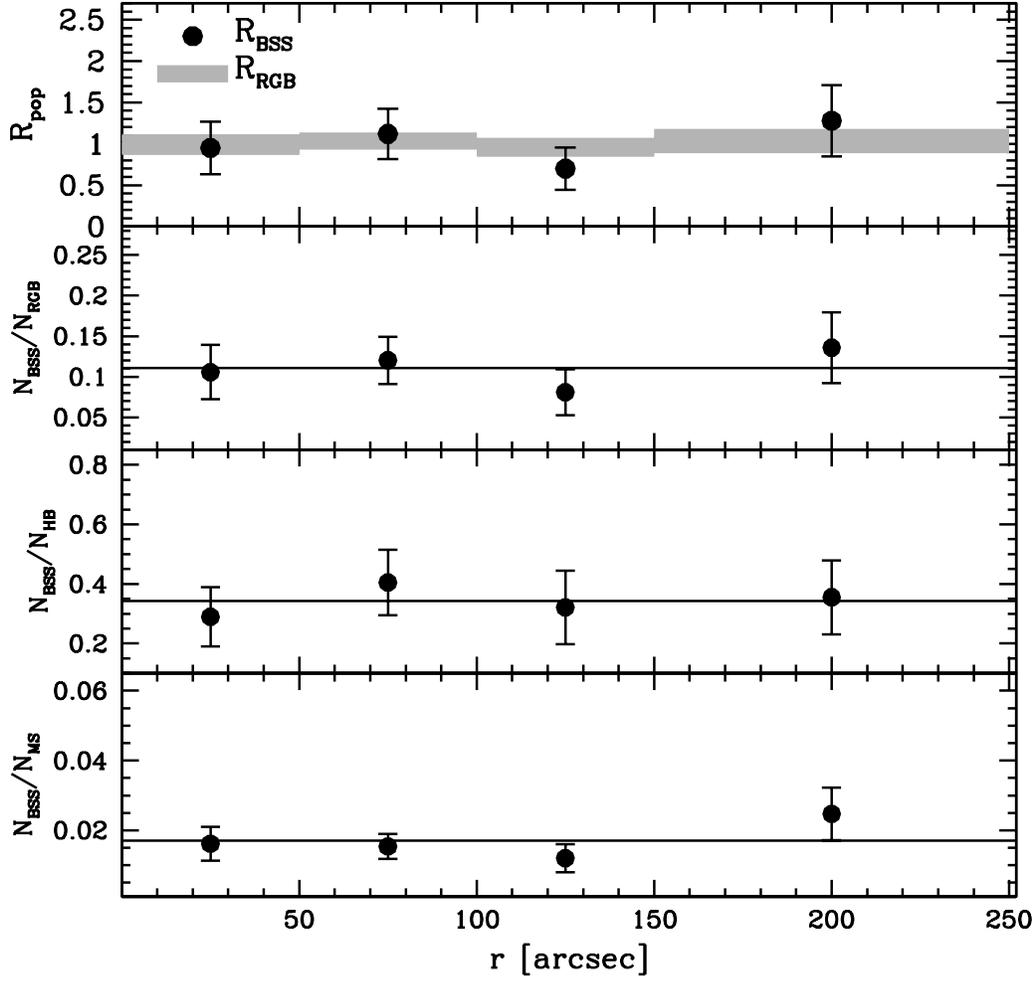}
\caption{From top to bottom, radial distributions of the double normalized ratios of the BSS 
($R_{\rm BSS}$; solid circles) and RGB stars ($R_{\rm RGB}$; grey regions), and the specific frequencies $N_{\rm BSS}/N_{\rm RGB}$, $N_{\rm BSS}/N_{\rm HB}$ and
$N_{\rm BSS}/N_{\rm MS}$ specific frequencies (solid circles).} 
\end{figure}

\begin{figure}
\includegraphics[scale=0.7]{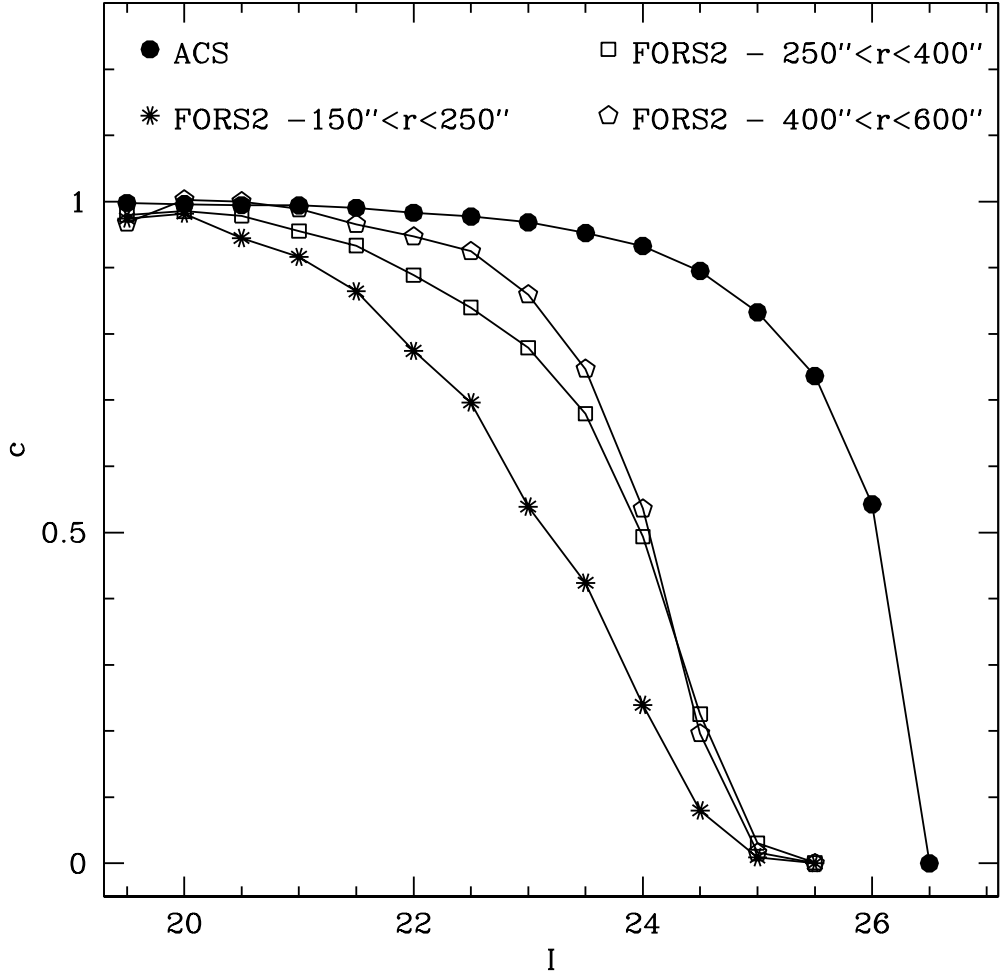}
\caption{Photometric completeness $C$ as a function of the $I$ band magnitude for the {\it ACS} and the {\it FORS2} data sets
and for different radial bins.}
\end{figure}

\begin{figure}
\includegraphics[scale=0.7]{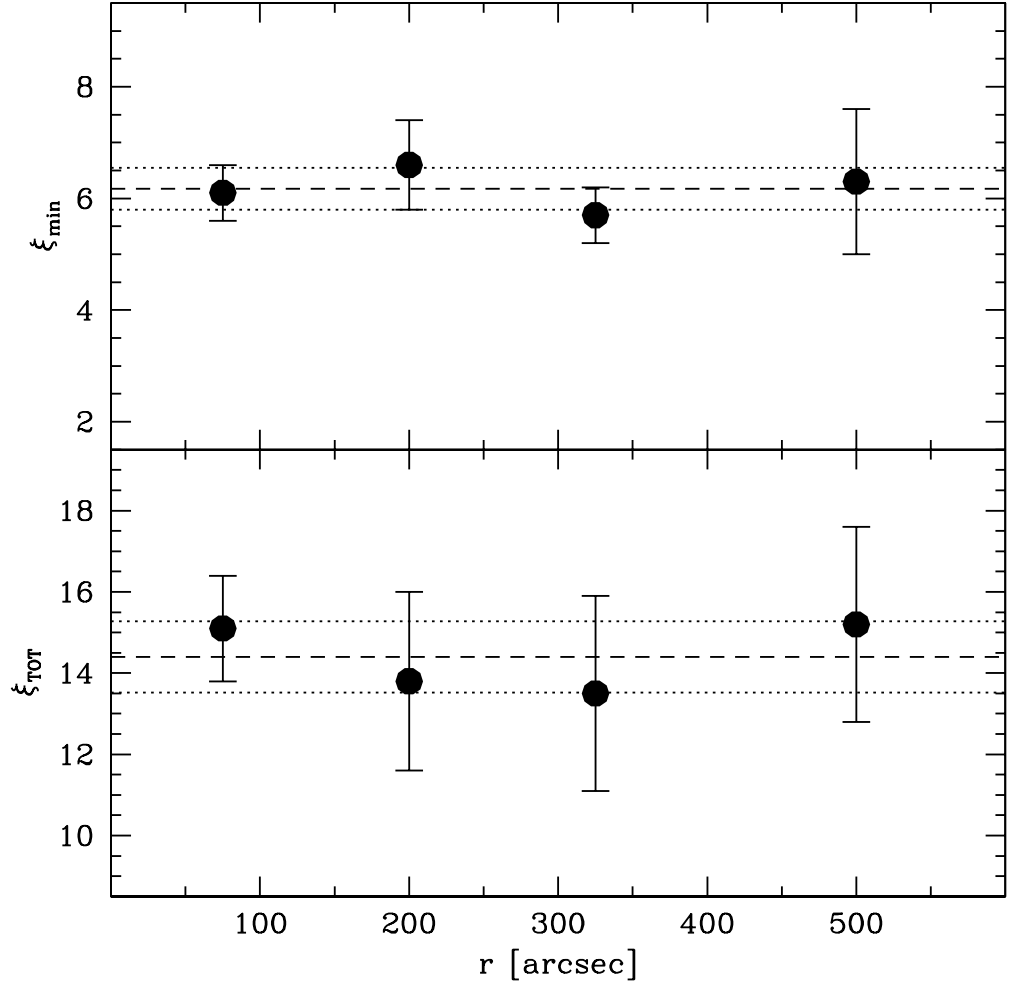}
\caption{Minimum and global binary fractions as a function of the radial distance from the cluster
center. The dashed lines represent the mean value, while the dotted ones define the $1 \sigma$ level.}
\end{figure}

\begin{figure}
\includegraphics[scale=0.7]{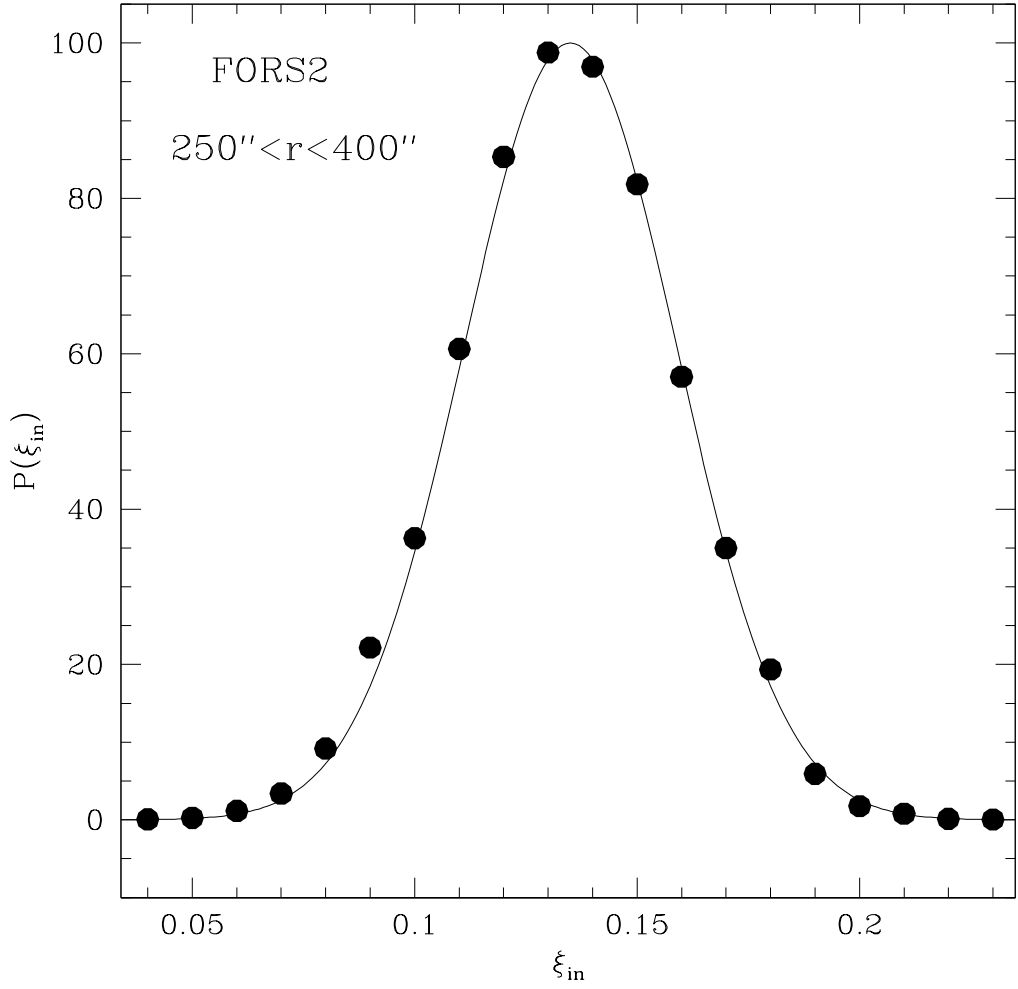}
\caption{Probability distribution of the adopted input binary fraction $\xi_{\rm in}$, for the case
of the FORS2 sample and for $250\arcsec<r<400\arcsec$}
\end{figure}

\begin{figure}
\includegraphics[scale=0.7]{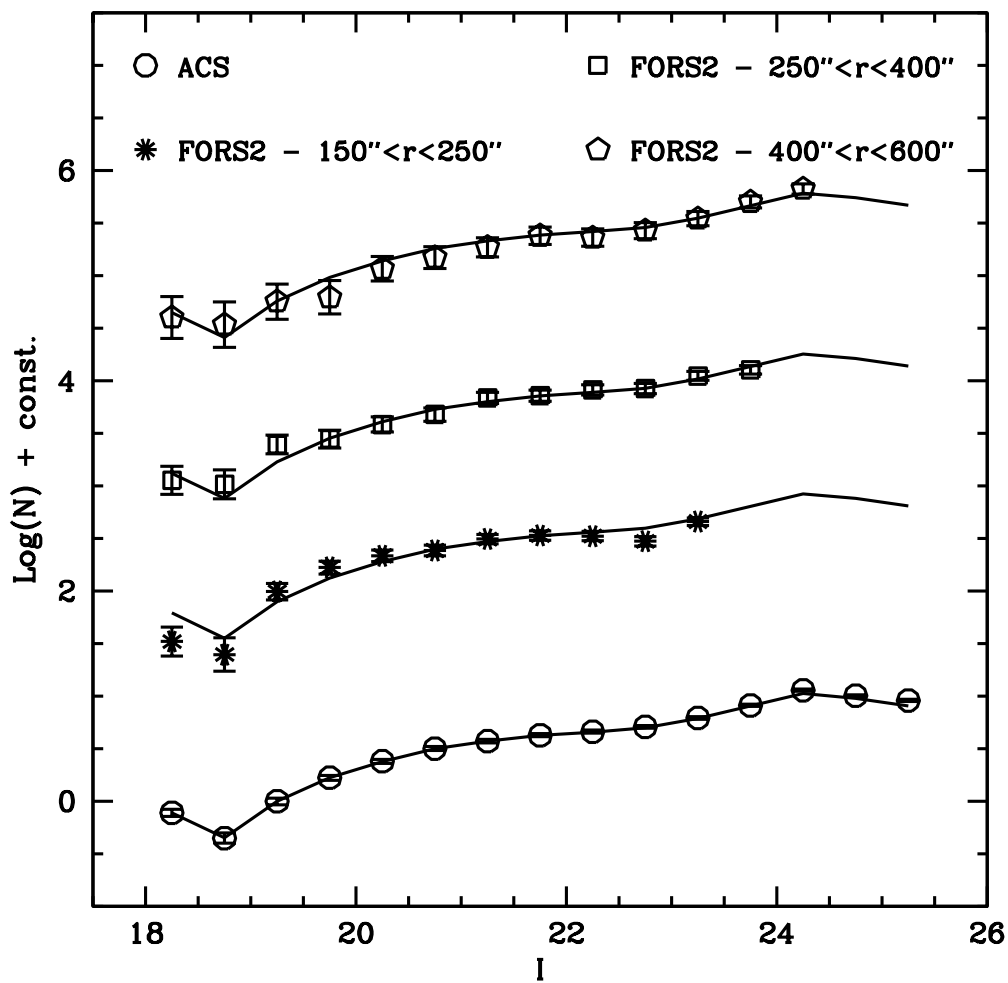}
\caption{LFs in the $I$ band obtained from the {\it ACS} and the {\it FORS2} data sets in different radial bins.
Measurements are shifted by an arbitrary amount to make the plot more readable. The solid line is the
LF obtained obtained in the ACS sample for $r<r_h$. It has been overplotted 
to the outer LFs for comparison. }
\end{figure}

\begin{figure}
\includegraphics[scale=0.7]{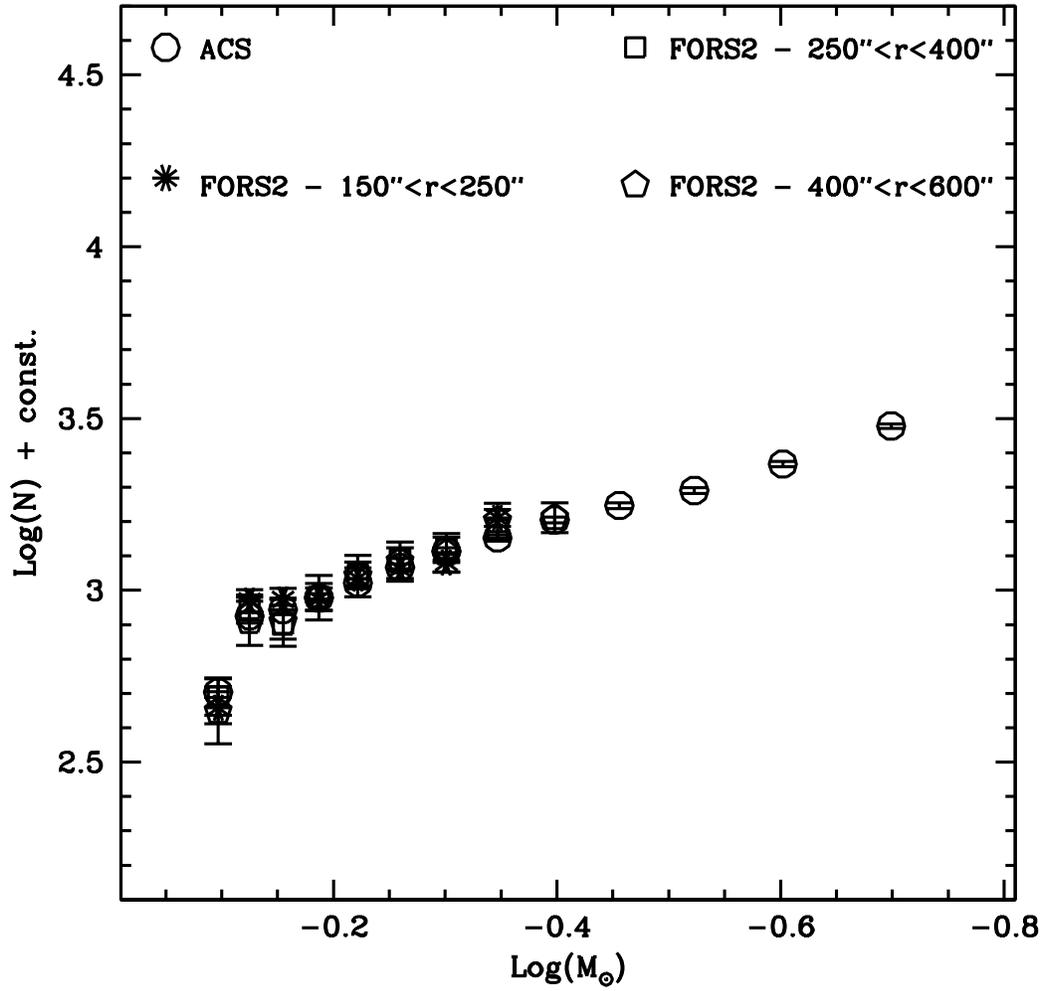}
\caption{MFs derived by using the Baraffe et al. (1997) mass-luminosity relation. 
Radial bins and symbols are the same as in Figure~12.}
\end{figure}
 
\end{document}